# Weak coupling theory of nickel-based 327 superconductors*

MIAO Jianjian[1,#], CHEN Weiqiang[1,2,†]

1. Quantum Science Center of Guangdong-Hong Kong-Macao Greater Bay Area, Shenzhen 518045, China
2. Department of Physics, Southern University of Science and Technology, Shenzhen 518055, China

**Abstract**

This review provides a comprehensive survey of weak-coupling theoretical approaches applied to understand the emergent superconductivity in the pressurized nickelate bilayer system $La_3Ni_2O_7$. Following the landmark discovery of its high-$T_c$ superconductivity under pressure, this material has rapidly become a new paradigmatic platform in the field of unconventional superconductivity, joining cuprates and iron-based systems. We focus on three pivotal theoretical frameworks: the random phase approximation (RPA), the fluctuation-exchange approximation (FLEX), and the functional renormalization group (FRG). These methods are deployed to analyze the effective pairing interactions and emergent instabilities arising from the low-energy electronic structure, which is commonly modeled by a bilayer two-orbital Hubbard Hamiltonian incorporating the Ni $d_{x^2-y^2}$ and $d_{3z^2-r^2}$ orbitals.

A central theme consolidating the weak-coupling perspective is the crucial role of Fermi surface topology and nesting. Theoretical studies consistently identify a multi-pocket Fermi surface under pressure, featuring an electron-like $\alpha$ pocket and two hole-like $\beta$ and $\gamma$ pockets. The $\gamma$ pocket, predominantly derived from the $d_{3z^2-r^2}$ orbital, exhibits strong nesting with the other pockets. This nesting significantly enhances antiferromagnetic spin fluctuations, which in turn mediate attractive pairing interactions. Consequently, a dominant $s_\pm$-wave pairing symmetry is widely predicted across different methodologies. In this state, the superconducting gap function maintains the same sign on the $\gamma$ and $\alpha$ pockets but reverses sign on the $\beta$ pocket, a structure intimately linked to the interlayer pairing channel dominated by the $d_{3z^2-r^2}$

---

orbital.

We further elaborate on the methodological distinctions and complementary strengths of the three weak-coupling approaches surveyed. The random phase approximation (RPA) provides an efficient description of spin and charge susceptibilities by summing an infinite series of bubble and ladder diagrams, offering a transparent link between Fermi surface nesting and the emergence of leading pairing instabilities. The fluctuation-exchange approximation (FLEX) improves upon RPA by incorporating self-consistency, thereby capturing the mutual renormalization between quasiparticle properties and collective spin fluctuations, which is essential for a more accurate determination of the pairing interaction in the intermediate coupling regime. The functional renormalization group (FRG) goes beyond static susceptibility calculations by integrating out high-energy degrees of freedom progressively, allowing for an unbiased treatment of competing instabilities—such as superconductivity, spin-density wave, and charge-density wave orders—within a unified framework. These complementary techniques collectively reinforce the conclusion that spin-fluctuation-mediated pairing, driven by the interlayer nesting between the $d_{3z^2-r^2}$-dominated γ pocket and other Fermi surface sheets, is the primary mechanism underlying superconductivity in the bilayer nickelate system.

The review systematically compares predictions from these approaches regarding the pairing strength, competition between different symmetries (e.g., s-wave vs. d-wave), and the influence of key material parameters such as interaction strength $U$, Hund's coupling $J_H$, crystal field splitting, and interlayer hybridization. Notably, the theoretical understanding is extended to ambient-pressure thin films, exploring the effects of substrate strain and the ongoing debate concerning the presence of the $\gamma$ pocket. We also discuss how the pressure dependence of $T_c$ and the material's response to disorder are addressed within the weak-coupling paradigm.

By synthesizing results from these advanced many-body techniques, this review highlights the success of weak-coupling, itinerant-based theories in capturing the essential physics of $La_3Ni_2O_7$, particularly in linking its high-$T_c$ superconductivity to spin-fluctuation-mediated pairing driven by specific Fermi surface geometries. Finally, we outline outstanding challenges and future directions, emphasizing the need for closer integration with strong-coupling pictures, more precise *ab initio*-derived model parameters, and definitive experimental tests to distinguish between competing theoretical scenarios and fully elucidate the superconducting mechanism in this fascinating nickelate system.

**Keywords:** Nickel-based high-temperature superconductivity; Random phase approximation; Fluctuation-exchange approximation; Functional renormalization group;

PACS: 74.70.-b; 71.10.-w; 74.20.-z; 05.10.Cc

# 1 Introduction

The electron pairing mechanism in high-temperature superconductors represents one of the most challenging and controversial topics in condensed matter physics. Theories of high-temperature superconductivity can be broadly categorized into strong-correlation and weak-coupling frameworks based on interaction strength. Strong-correlation theories typically address scenarios involving strong Coulomb interactions. A representative example is the doped Mott insulator[1], while the Hubbard model (in the large-U limit) and the *t*-*J* model[2] serve as common strong-correlation models. In contrast, weak-coupling theories originate from Landau Fermi liquid theory. These approaches investigate superconducting mechanisms by analyzing effective interactions among quasiparticles near the Fermi surface and assessing Fermi surface instabilities. The theory of exchange antiferromagnetic spin fluctuations[3] is a prominent example of weak-coupling theory. However, many high-temperature superconductivity theories do not fit neatly into this binary classification. For instance, phenomenological theories explain phenomena such as strange metals[4] and pseudogaps in the underdoped region[5] of cuprate high-temperature superconductors. The discovery of nickel-based 112 superconductors[6] in 2019, followed by the achievement of a superconducting transition temperature near 80 K in nickel-based 327 superconductors[7] under high pressure in 2023, has sparked a third wave of research interest in high-temperature superconductivity, following the earlier waves focused on cuprates and iron-based materials. This article summarizes recent progress in weak-coupling theories related to nickel-based 327 superconductors. It covers studies on superconducting pairing mechanisms and symmetry using theoretical methods such as the random phase approximation, fluctuation-exchange approximation, and functional renormalization group. For the convenience of readers, we also provide brief introductions to these theoretical methods in the main text.

# 2 Multi-orbital Hubbard Model

The theoretical starting point for $La_3Ni_2O_7$ is the formal valence state of nickel (Ni) as $2.5+$ (corresponding to the $3d^{7.5}$ electron configuration). This directly implies a multi-orbital electronic structure. A simplified picture considering crystal field splitting and interlayer coupling indicates that the Ni $t_{2g}$ orbitals and the bonding $3d_{z^2}$ state are fully occupied. This leaves the $3d_{x^2-y^2}$ orbital at quarter-filling[7], while the antibonding $3d_{z^2}$ orbital remains empty. This description is qualitatively consistent with density functional theory (DFT) calculations. DFT identifies three key bands near the Fermi level: two derived from Ni $3d_{x^2-y^2}$ orbitals ($\alpha$ and $\beta$ bands) and one from the bonding Ni $3d_{z^2}$ state ($\gamma$ band), as shown in Fig. 1[8-13]. The position of this $\gamma$ band relative to the Fermi level has become a central point of current debate. In non-superconducting bulk samples at ambient pressure, both DFT and angle-resolved photoemission spectroscopy (ARPES) consistently find that the $\gamma$ band lies entirely below the Fermi level. Consequently, only the $\alpha$ and $\beta$ bands constitute the Fermi surface[14-16]. However, the theoretical picture for the superconducting state is more controversial. Under pressure, many DFT calculations predict that structural changes induce a hole pocket derived from the $\gamma$ band. Some researchers initially considered this feature crucial for achieving high $T_c$[7,13,17-19]. Nevertheless, the existence of the $\gamma$ pocket in 327 thin films at ambient pressure remains debated, both experimentally[20-22] and theoretically[23-27].

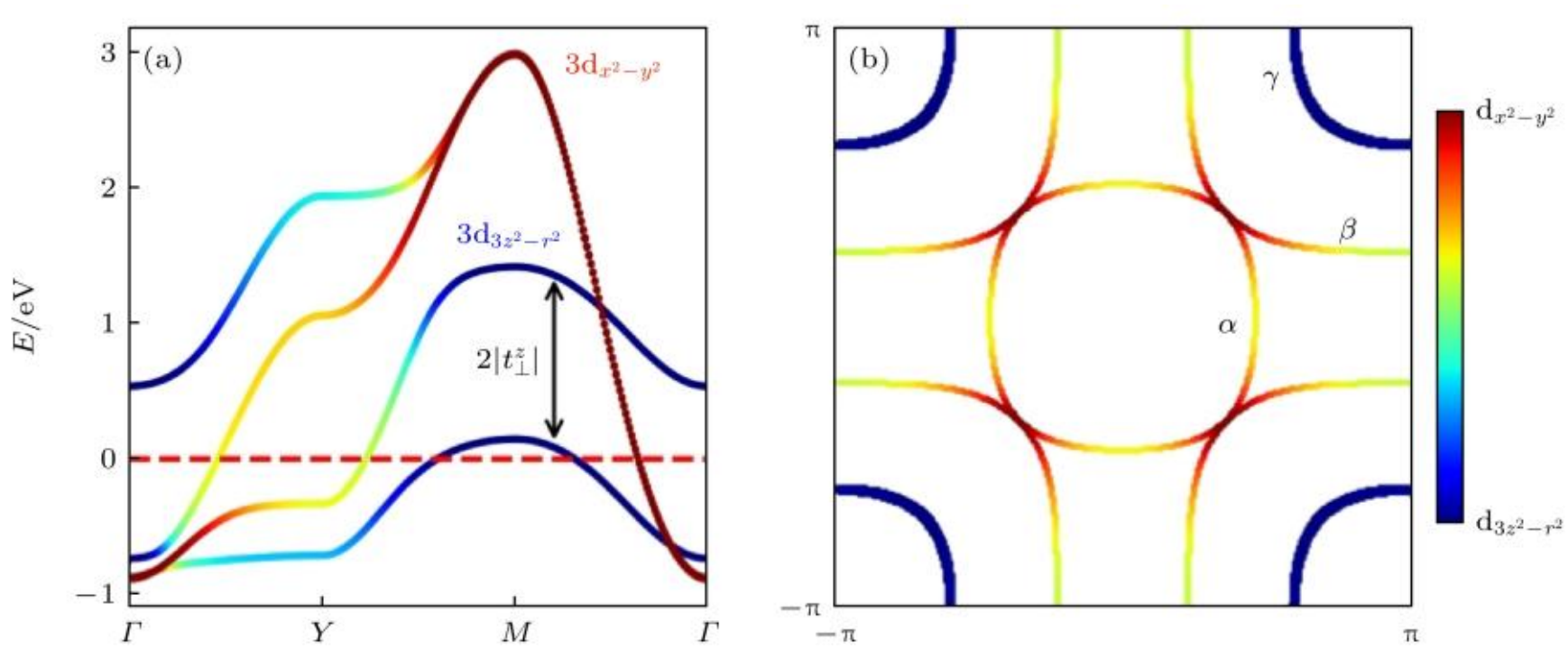

**Fig. 1 (a) The band structure and (b) the Fermi surface of the bilayer two-orbital**

**model[8]. The colored bar indicates the orbital weight of the two orbitals $d_{x^2-y^2}$ and $d_{3z^2-r^2}$.**

Nevertheless, most researchers agree that the electronic transport in nickel-based 327 materials primarily originates from the nickel $3d_{x^2-y^2}$ and $3d_{3z^2-r^2}$ orbitals. Within the weak-coupling theory, the microscopic description of nickel-based 327 superconductors typically employs a two-orbital Hubbard model constructed on a bilayer square lattice. This model comprises two components: a non-interacting tight-binding model and multi-orbital Hubbard interactions. The Hamiltonian for the tight-binding part is expressed as

$$H_0 = \sum_{ij,\mu\nu,\sigma} t_{ij}^{\mu\nu} c_{i\mu\sigma}^\dagger c_{j\nu\sigma}, \tag{1}$$

Here, $i/j$ denotes lattice sites, $\sigma$ denotes spin, and $\mu/\nu$ denotes orbitals ($d_{x^2-y^2}$ or $d_{3z^2-r^2}$). The hopping integrals $t_{ij}^{\mu\nu}$ can be obtained via density functional theory calculations[8]. Figure 1 presents the results of the density functional calculations. Under high pressure, the Fermi level intersects only three of the four bands, thereby generating three Fermi pockets. Among these, the pocket labeled $\alpha$ is an electron pocket, whereas the $\beta$ and $\gamma$ pockets are hole pockets. Notably, the $\alpha$ and $\beta$ pockets lie in close proximity. These two pockets exhibit significant hybridization between two orbitals, while the $\gamma$ pocket consists primarily of the $d_{3z^2-r^2}$ orbital. Importantly, DFT calculations indicate that the $\gamma$ pocket emerges only under pressure, which aligns with the presence of superconductivity under pressure. This suggests that the $\gamma$ pocket may play a critical role in the superconductivity of the system.

The Hubbard interaction term is generally expressed as

$$\begin{aligned} H_{\mathrm{I}} = \ & U \sum_{i\mu} n_{i\mu\uparrow} n_{i\mu\downarrow} + V \sum_{i\sigma\sigma'} n_{i1\sigma} n_{i2\sigma'} \\ & + J_{\mathrm{H}} [\sum_{i\sigma\sigma'} c_{i1\sigma}^\dagger c_{i2\sigma'}^\dagger c_{i1\sigma'} c_{i2\sigma} \\ & + \sum_i c_{i1\uparrow}^\dagger c_{i1\downarrow}^\dagger c_{i2\downarrow} c_{i2\uparrow} + \mathrm{H.c.}], \end{aligned} \tag{2}$$

The term $U(V)$ represents the intra-orbital (inter-orbital) Hubbard repulsion, while $J_{\mathrm{H}}$ denotes the Hund's coupling and pair hopping. Generally, the relation $U = V + 2J_{\mathrm{H}}$ holds. When the interaction parameter $U$ is not excessively large, this model can be treated using weak-coupling methods. The following section primarily introduces studies on the superconducting pairing mechanism and pairing symmetry within this model, based on theoretical approaches such as the random phase approximation, fluctuation exchange approximation, and functional renormalization group.

# 3 Random Phase Approximation

The random phase approximation (RPA) was originally developed by Bohm and Pines[28] as a quantum many-body method for treating degenerate electron gases with long-range Coulomb interactions. This approach reveals that the effective interaction between electrons is a short-range screened Coulomb interaction. In more modern quantum many-body theory, the random phase approximation is equivalent to the large-$N$ limit. Its core idea involves generalizing electrons with two spin degrees of freedom in a degenerate electron gas to fermions with $N$ degrees of freedom. Consequently, the corresponding symmetry extends from $SU(2)$ to $SU(N)$. When $N$ is large, contributions of different orders can be distinguished via a power series expansion in $1/N$. The results of the random phase approximation are obtained in the limit where $N \to \infty$. Taking the calculation of the effective interaction in a degenerate electron gas as an example, Feynman diagram techniques can represent different contributions using distinct diagrams. As shown in Figure 2[29], the effective interaction, represented by double wavy lines, can be expressed as an infinite geometric series expansion. Here, single wavy lines represent the unscreened long-range Coulomb interaction, with each single wavy line contributing a factor of $1/N$. Various other complex contributions are encapsulated within the polarization bubble labeled $\chi$, leaving the calculation of this polarization bubble as the central challenge. The polarization bubble can also be expanded using Feynman diagrams, as illustrated in Figure 3[29]. Contributions of different orders are distinguished by the number of single wavy lines included in the diagrams, where solid lines with arrows represent the electron Green's functions. If only the lowest-order contribution is considered—specifically, the bubble formed by connecting the lines corresponding to the Green's functions of two electrons end-to-end—and higher-order contributions are ignored, summing the previous infinite geometric series in this limit yields the same effective interaction as that derived from the random phase approximation. In real space, this interaction exhibits an exponentially decaying form, corresponding to a short-range screened Coulomb interaction.

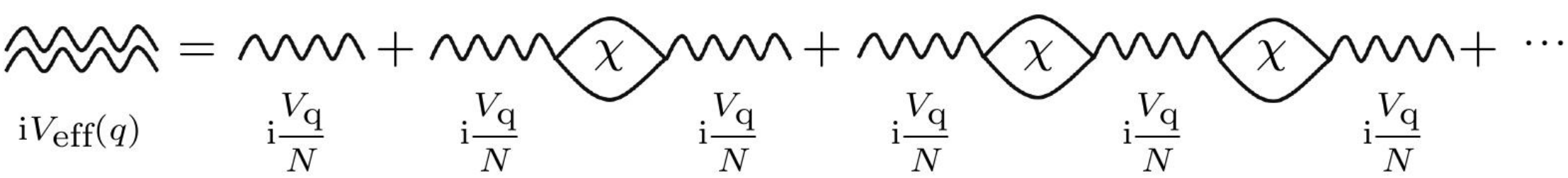

**Fig. 2 Feynman diagram expansion of the effective interaction. The single wavy**

**lines represent the unscreened long-range Coulomb interaction, and the double wavy line represents the effective interaction.**

$\chi \; = \; + \; + \; + \; + \cdots = \mathrm{i}N\chi(q)$

$O(N) \quad O(1) \quad O(1) \quad O(1/N)$

**Fig. 3 Feynman diagram expansion of the polarization bubble, where solid lines with arrows represent the electron Green's function.**

The random phase approximation can also be applied to investigate magnetic susceptibility and charge susceptibility in interacting systems, which describe spin and charge density fluctuations, respectively. In the non-interacting case, these quantities are represented in Feynman diagrams by bubble loops formed by connecting two electron Green’s function lines end-to-end. These non-interacting susceptibilities and polarizabilities can be calculated rigorously, and they are all equal. For interacting systems, obtaining exact results for different susceptibilities and polarizabilities is generally difficult. However, they can be calculated using Feynman diagram techniques within the framework of perturbation theory. Under the random phase approximation, only the end-to-end connected bubble loops are retained. This simplifies the calculation of magnetic susceptibility and charge susceptibility to summing an infinite geometric series, as shown in Figure 4[29]. By converting the infinite series summation into a Dyson equation, one can solve for the magnetic susceptibility and charge susceptibility under the random phase approximation.

**Fig. 4 Feynman diagram expansion of the susceptibility at the RPA level. The second line shows the Feynman diagram representation of the Dyson equation.**

Charge susceptibility and magnetic susceptibility allow for the investigation of system instabilities arising from charge density and spin density fluctuations. As interactions gradually increase, susceptibility diverges rapidly at specific wave vectors. This divergence indicates a tendency toward charge or spin ordered states with specific wave vectors, namely charge density wave (CDW) or spin density wave (SDW) order. The interaction strength at which these fluctuations diverge is termed the critical interaction strength, denoted as $U_{\mathrm{c}}^{\mathrm{CDW}}$ and $U_{\mathrm{c}}^{\mathrm{SDW}}$. By comparing the rates at which electric and magnetic susceptibilities diverge, or by comparing the magnitudes of the corresponding $U_{\mathrm{c}}$, one can determine whether the system favors CDW or SDW formation and identify the specific wave vector. For multi-orbital Hubbard models with repulsive interactions ($U > 0$), $U_{\mathrm{c}}^{\mathrm{CDW}} > U_{\mathrm{c}}^{\mathrm{SDW}}$ generally holds. This implies that SDW order typically forms more readily. When the interaction $U$ exceeds the critical interaction $U_{\mathrm{c}}$, perturbation theory fails, rendering results from the RPA unreliable. Consequently, the applicability of RPA is limited to regimes where neither electric nor magnetic susceptibility has fully diverged. Low-energy scattering processes near the Fermi surface dominate magnetic susceptibility. Therefore, susceptibility magnitude depends heavily on Fermi surface structure. The specific wave vector in SDW often corresponds to the nesting vector of a Fermi surface with nesting structures.

An early application of RPA in superconductivity was the study by Berk and Schrieffer[30] on the suppression of conventional s-wave spin-singlet superconductivity by ferromagnetic spin fluctuations. Subsequently, Anderson and Brinkman[31] and Nakajima[32] recognized in their studies of helium-3 superfluidity that spin fluctuations not only stabilize spin-triplet p-wave states but also serve as a key pairing mechanism. It is well established that conventional superconductors in metals and alloys are explained by Bardeen-Cooper-Schrieffer (BCS) theory. In this framework, effective attractive interactions between electrons, mediated by phonon exchange, overcome screened Coulomb repulsion near the Fermi surface. This leads to Cooper pair formation, Fermi surface instability, and coherent condensation into a superconducting state. However, the discovery of unconventional superconductors, such as heavy fermion, organic, and cuprate superconductors, revealed discrepancies between experimental data and BCS theory based on electron-phonon coupling. Hirsh and others proposed that effective attractive interactions could arise from the exchange of antiferromagnetic or SDW fluctuations, leading to pairing and superconductivity[33-35].

The fundamental physical picture stems from strong spin fluctuations near the SDW transition point. Here, electrons near the Fermi surface generate effective pairing interactions through spin fluctuation exchange, forming Cooper pairs. Perturbation theory allows for the quantitative calculation of these effective pairing interactions. Within RPA, corrections to the effective pairing interaction arise from two types of diagrams: bubble diagrams and ladder diagrams, as shown in Figure 5[36]. Summing these diagrams is analogous to summing infinite geometric series in screened Coulomb interaction calculations. The corrections to the effective pairing interaction correspond to electric and magnetic susceptibilities under RPA. Quantitative calculations account for both charge density and spin fluctuations. However, since magnetic susceptibility typically diverges faster, spin fluctuations provide the dominant contribution. As temperature decreases, Cooper pairs undergo coherent condensation, driving the system into a superconducting state. Using the effective pairing interaction combined with mean-field approximation yields a self-consistent equation for the superconducting gap function. Near the critical superconducting temperature $T_c$, this equation simplifies to a linear form. This transforms the solution of the gap function into an eigenvalue problem for the effective pairing interaction matrix. The eigenfunction corresponding to the largest eigenvalue represents the dominant superconducting gap function, allowing determination of the pairing symmetry.

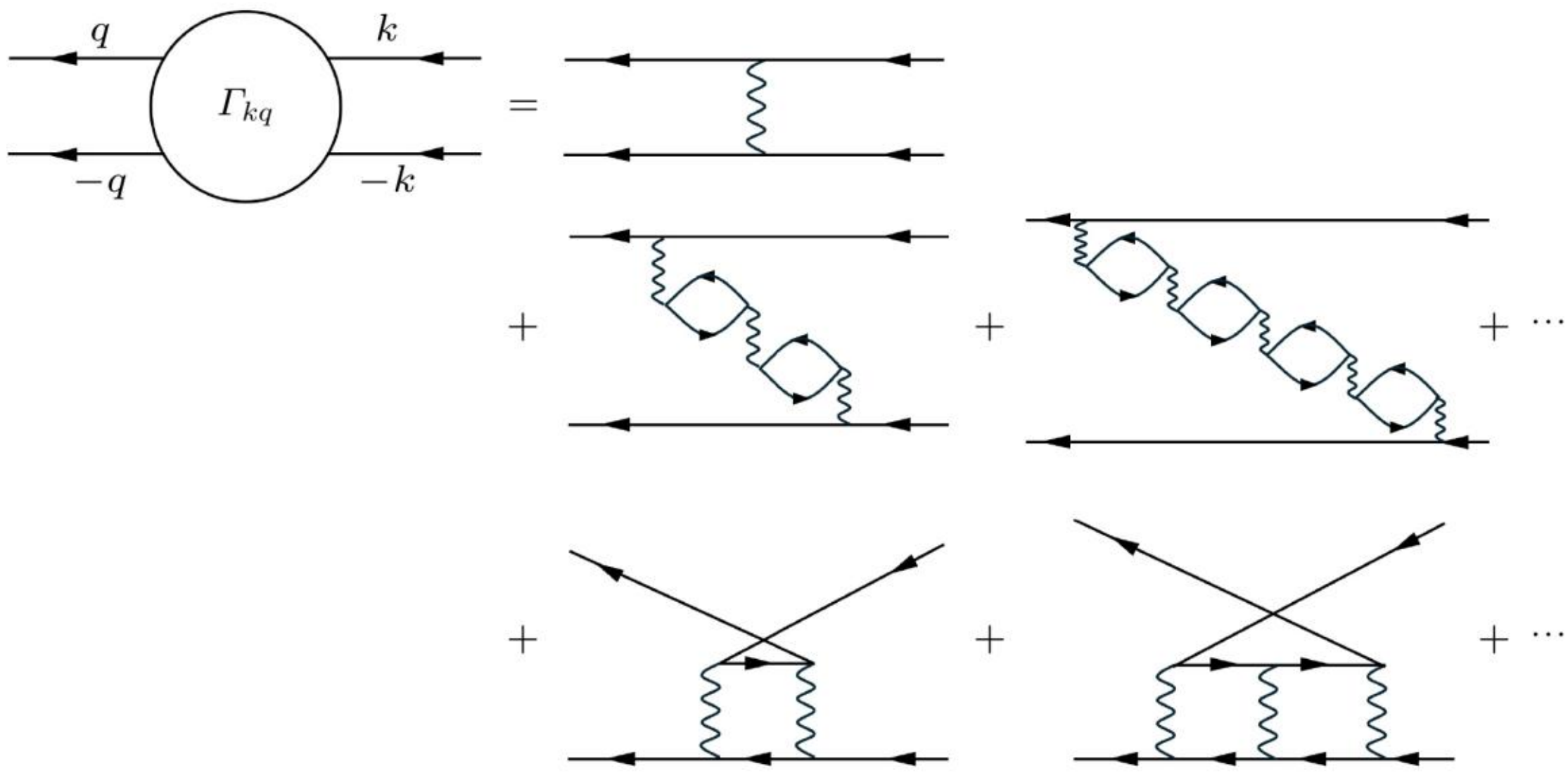

**Fig. 5 Feynman diagram expansion of the effective pairing interaction $\Gamma_{kq}$ at the RPA level. The second line includes bubble contributions, and the third line includes ladder contributions.**

Weak-coupling theories based on the random phase approximation have been applied in studies of both copper-based and iron-based high-temperature superconducting materials. Prior to the experimental confirmation of d-wave pairing symmetry in copper-based superconductors[37,38], Monthoux et al.[39] predicted that these materials should exhibit $d_{x^2-y^2}$-wave symmetry. This prediction was based on a pairing mechanism driven by antiferromagnetic fluctuations exchanged between electrons. Scalapino et al.[40] performed calculations on a single-band Hubbard model on a three-dimensional cubic lattice using the random phase approximation. They found that the system tends to form unconventional d-wave pairing near the spin density wave instability. Kuroki et al.[41] constructed a five-band model for iron-based superconductors, incorporating all d orbitals. By investigating superconductivity using the random phase approximation, they discovered that multiple spin fluctuations induced by nesting structures between disconnected Fermi surfaces cause the system to favor extended s-wave pairing. Graser et al.[42] employed the random phase approximation to study pairing interactions and pairing instabilities in the five-orbital model of iron-based superconductors. They found that due to the near-nesting between multiple Fermi surface sheets, the two dominant pairing symmetries in iron-based superconductors, $s\pm$-wave and $d_{x^2-y^2}$-wave, are nearly degenerate.

The random phase approximation method has been widely applied in theoretical studies of nickel-based 327 superconductors. Yao Daoxin et al.[8] were the first to use density functional theory to derive a bilayer two-orbital model and its tight-binding parameters for nickel-based 327 superconductors under high pressure. They pointed out that the Fermi surface consists of two electron pockets, $\alpha$ and $\beta$, and one hole pocket, $\gamma$, as shown in Figure 1. By calculating the spin susceptibility using the random phase approximation, they found that the main contribution arises from scattering between $d_{3z^2-r^2}$ orbitals. This reflects the nested structure of the Fermi surface of the $\gamma$ pocket formed by $d_{3z^2-r^2}$ orbitals. Similar spin susceptibility results have been obtained in subsequent studies[10]. Yao Daoxin et al.[43] also used the same theoretical framework to further investigate the pairing mechanism of nickel-based 327 superconducting thin films at ambient pressure. Yang Fan, Chen Weiqiang, and others[18] conducted a more systematic study of nickel-based 327 superconductors under high pressure using the random phase approximation. By calculating the charge susceptibility and magnetic susceptibility of the bilayer two-orbital model within the random phase approximation, they found three sets of inequivalent peaks in the distribution of the maximum eigenvalue of the spin susceptibility matrix across the Brillouin zone. These are denoted as $Q_1$, $Q_2$, and $Q_3$, as shown in Figure 6. Among these, the $Q_1$ peak is the largest, corresponding to the nesting between the $\gamma$ and $\beta$ pockets. When the interaction strength is around 1 eV, there is significant competition between s-wave and d-wave pairing. However, as the interaction increases further, s-wave pairing ultimately becomes dominant. By analyzing the distribution of the gap function near the Fermi surface, it can be observed that the s-wave gap function has the same sign on the $\gamma$ and $\alpha$ pockets, but the opposite sign on the $\beta$ pocket. This indicates $s_\pm$-wave pairing, as shown in Figure 7. This result suggests that the $\gamma$ pocket plays a key role in forming superconductivity. Furthermore, real-space pairing is dominated by interlayer $d_{3z^2-r^2}$ orbitals. Given that the $\gamma$ pocket is primarily composed of $d_{3z^2-r^2}$ orbitals, this further validates the importance of the $\gamma$ pocket. Considering that apical oxygen defects are easily generated in actual materials, they also used real-space random phase approximation to study the impact of apical oxygen defects on nickel-based 327 superconductors under high pressure. They found that these defects not only reduce the average interlayer electron hopping but also generate local magnetic moments near the defects, thereby significantly suppressing superconductivity. They also pointed out that preparing materials in an oxygen-rich environment can effectively increase the superconducting critical temperature $T_c$.

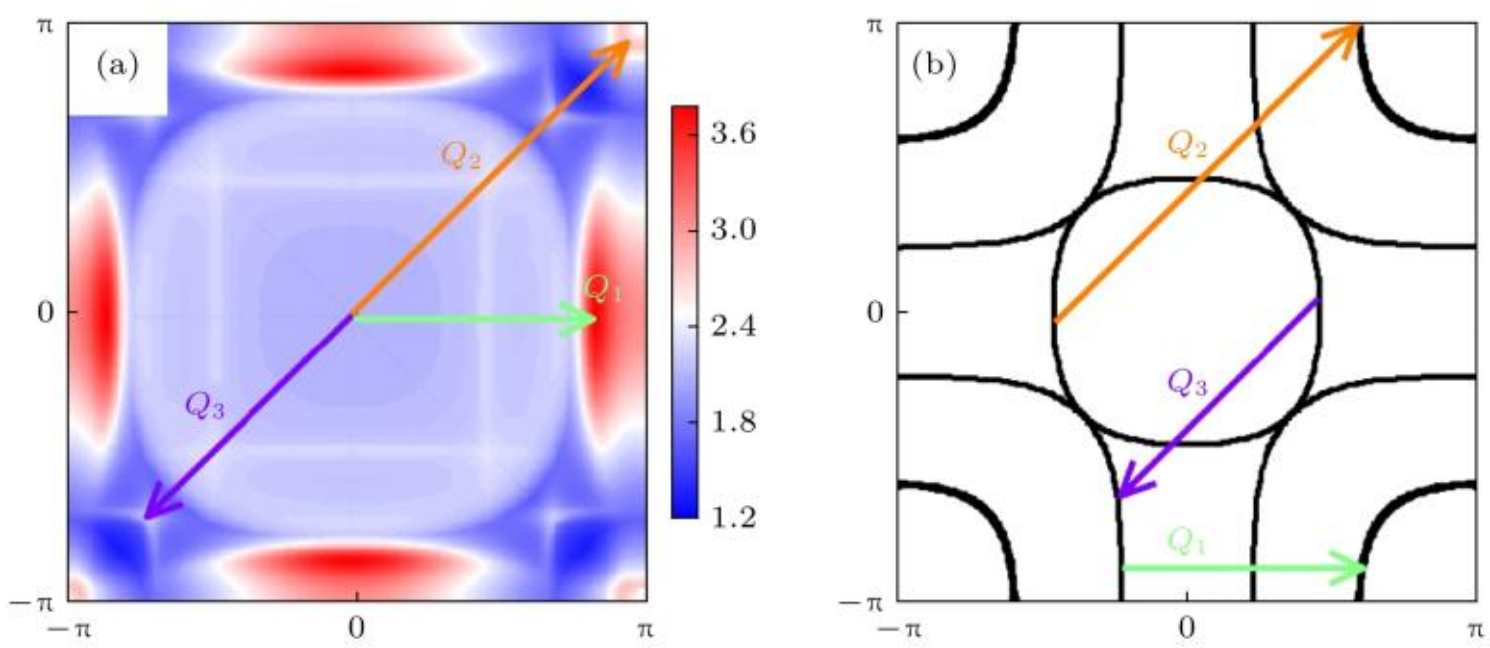

**Fig. 6 (a) Distribution of the largest eigenvalue $\chi(q)$ of the RPA-renormalized spin susceptibility matrix within the Brillouin zone[18]. This distribution exhibits peaks at three inequivalent wavevectors, denoted as $Q_1$, $Q_2$, and $Q_3$, respectively; (b) Fermi surface nesting of the *α*, *β*, and *γ* pockets, with the nesting wavevectors $Q_1$, $Q_2$, and $Q_3$ indicated[18].**

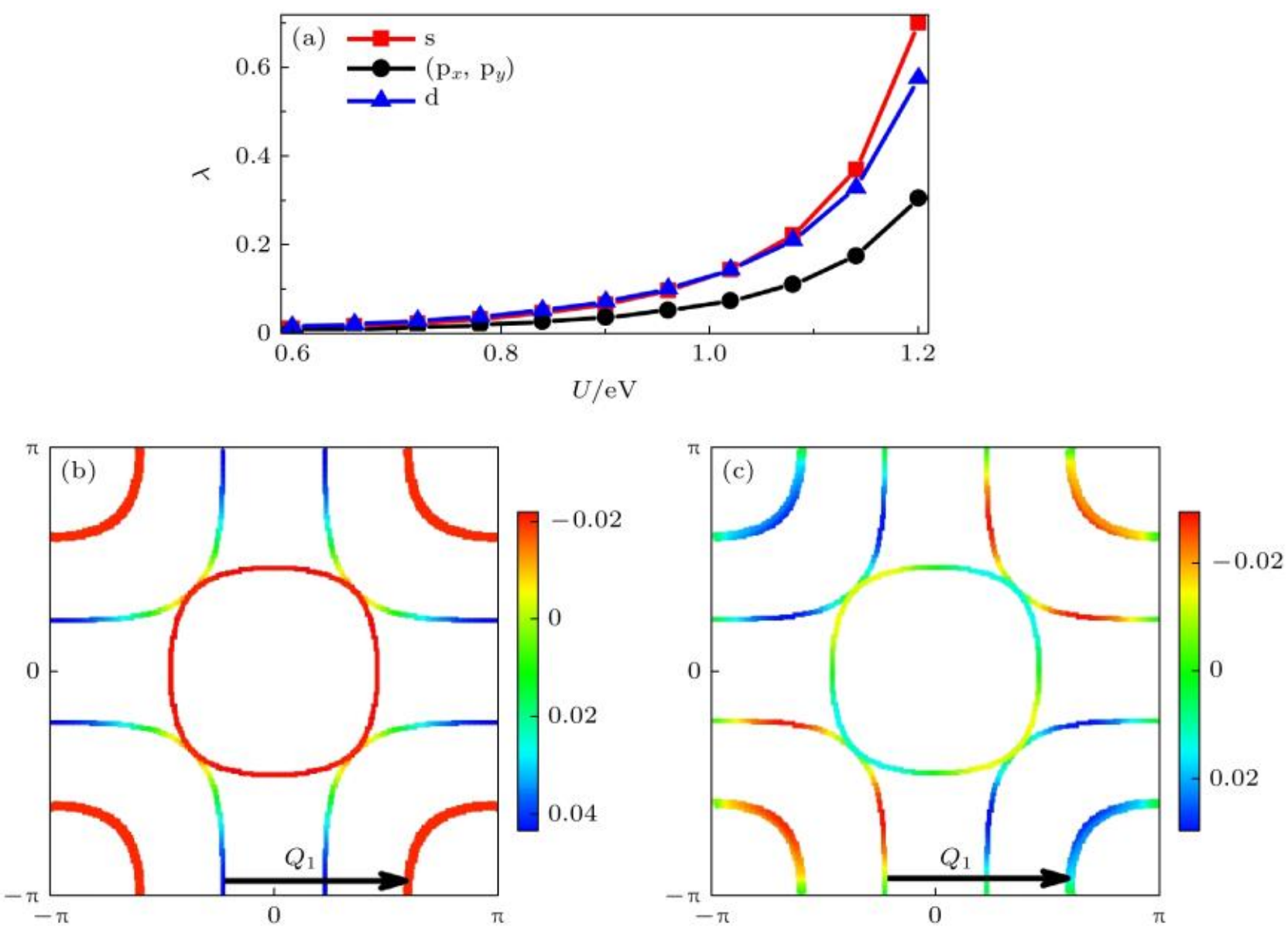

**Fig. 7 (a) The largest eigenvalue $\lambda$ of the linearized gap equation for various pairing symmetries as a function of the interaction strength $U$ with fixed $J_{\mathrm{H}} = U/6$; (b), (c) the distributions of the leading s- and d-wave pairing gap functions on the Fermi surface for $U = 1.16$ eV[18].**

By investigating the effective pairing interaction within the random phase approximation and solving the corresponding eigenvalue problem for the gap function, Dagotto et al.[44] also found that the dominant pairing symmetry in nickel-based 327 superconductors is $s_{\pm}$-wave pairing. The pairing mechanism primarily arises from the nesting structure between the Fermi surface pocket centered at $M = (\pi,\pi)$ and those centered at $X = (\pi, 0)$ and $Y = (0,\pi)$. This finding aligns with the $s_{\pm}$-wave pairing mechanism induced by nesting between the $\gamma$ pocket and the $\beta$ and $\alpha$ pockets, as reported by Yang Fan, Chen Weiqiang, and colleagues. In subsequent work, Dagotto et al.[45] systematically studied nickel-based 327-like superconductors theoretically by substituting lanthanum with other rare-earth elements. They found that for nickel-based 327-like superconductors with a stable *Fmmm* phase, the dominant pairing symmetry remains $s_{\pm}$-wave pairing. The critical pressure increases with the atomic number, while the superconducting critical temperature decreases as the atomic radius diminishes. Considering that the atomic radius of lanthanides gradually decreases from lanthanum to lutetium as the atomic number increases, the lattice constants of the resulting nickel-based 327-like superconductors also decrease. This reduction enhances both intralayer and interlayer hopping, effectively reducing $U/W$, where $W$ represents the bandwidth. These results indicate that appropriately enhancing electronic correlations in the system can promote superconductivity. Since nickel-based 327 superconductors already contain the first lanthanide element, Dagotto et al. suggested that the superconducting critical temperature $T_c$ could be increased by growing nickel-based 327 superconductors on substrates with larger in-plane lattice constants.

Chen Hanghui et al.[46] investigated the influence of crystal field splitting on nickel-based 327 superconductors under high pressure and found that their pairing symmetry is highly sensitive to the magnitude of crystal field splitting. Based on the band structure obtained from density functional theory and maximally localized Wannier functions, the superconducting pairing symmetry derived using the random phase approximation is $d_{xy}$-wave pairing. However, a slight increase in crystal field splitting causes the pairing symmetry of the system to transition from $d_{xy}$-wave to $s_{\pm}$-wave pairing. Bötzel et al.[47] studied the magnetic excitations of nickel-based 327 superconductors under high pressure and found that the pairing symmetry of the system can be effectively determined through the spin response of the bilayer structure. For interlayer-driven $s_{\pm}$-wave pairing, a large spin resonance peak appears near the *X* point only in the odd channel. In contrast, for d-wave pairing, spin resonances are present in both the odd and even channels.

The discovery of nickel-based 327 superconductor thin films at ambient pressure[48,49] has enabled the application of more experimental measurement techniques to these films, laying the foundation for further elucidation of the superconducting mechanism and pairing symmetry. Chen Weiqiang, Yue Changming, and colleagues[20] combined density functional theory, cluster dynamical mean-field theory, and constrained random phase approximation to obtain Fermi surfaces consistent with angle-resolved photoemission spectroscopy (ARPES) measurements. Unlike the Fermi surfaces derived from simple density functional theory, their results reveal that correlation effects play an indispensable role in nickel-based 327 superconductors. Based on this, calculations using the random phase approximation revealed that the dominant pairing instability is $s_{\pm}$-wave pairing. This pairing originates from strong spin fluctuations caused by Fermi surface nesting, which primarily occurs in bands composed of $d_{3z^2-r^2}$ orbitals[23]. However, other ARPES experiments have found that the nickel $3d_{z^2}$ bonding band lies below the Fermi level, resulting in the absence of the $\gamma$ pocket in the Fermi surface of nickel-based 327 superconductor thin films at ambient pressure[21,22]. Based on the band structure without the $\gamma$ pocket provided by DFT, Yang Fan, Wu Congjun, and colleagues[24] used the random phase approximation to calculate $s_{\pm}$-wave pairing induced by nesting between the $\alpha$ and $\beta$ pockets. In real space, this corresponds to interlayer $d_{x^2-y^2}$ pairing. In the strong coupling limit, slave-boson mean-field calculations on a two-orbital *t-J* model revealed that interlayer superexchange induced by Hund's coupling between $d_{z^2}$ orbitals also leads to interlayer s-wave pairing for $d_{x^2-y^2}$, consistent with conclusions drawn in the weak coupling limit.

# 4 Fluctuation-Exchange Approximation

The fluctuation-exchange approximation (FLEX) is a quantum many-body method originally proposed by Bickers et al.[50,51] for handling strongly correlated electron systems. FLEX goes beyond simple mean-field theories, such as the Hartree-Fock approximation, and can handle two-particle correlation functions with strong frequency and momentum dependence. More importantly, FLEX satisfies microscopic conservation laws and sum rules in its calculations. Consequently, it was initially referred to as a conserving approximation. Its core concept can be traced back to the approximate solution methods for multi-particle Green's functions developed by Baym and Kadanoff to ensure that transport results satisfy conservation laws[52,53]. Similar to the random phase approximation, which considers charge density and spin fluctuations[54], the Feynman diagrams for the self-energy function and effective pairing interaction in FLEX also include bubble and ladder diagrams, as shown in Figures 8 and 9[55]. The difference lies in the representation of the solid lines with arrows in the Feynman diagrams. In the random phase approximation, these lines represent "bare" single-particle Green's functions, allowing for direct summation of infinite geometric series. In contrast, in FLEX, these lines represent "dressed" single-particle Green's functions, requiring self-consistent solutions for the infinite series. The self-consistent cycle for solving Green's functions and susceptibility using FLEX is as follows: First, calculate the non-interacting single-particle Green's function $G^{(0)}$. Determine the initial chemical potential based on particle number using $G^{(0)}$. Combine the "bare" interaction $V$ and $G^{(0)}$ to provide an initial self-energy $\Sigma^{(0)}$. Then, obtain the "dressed" single-particle Green's function $G$ via the Dyson equation. Use $G$ to obtain the susceptibility $\chi^{(0)}$. Next, calculate the "dressed" susceptibility $\chi^{\mathrm{RPA}}$ under the random phase approximation and use $\chi^{\mathrm{RPA}}$ to derive the effective interaction $V_{\mathrm{eff}}$. Combine $V_{\mathrm{eff}}$ and $G$ to calculate the new self-energy $\Sigma$. Finally, obtain the new "dressed" single-particle Green's function $G$ again using the Dyson equation, completing one cycle. By iterating this cycle until convergence, the Green's function and susceptibility under FLEX are obtained. The flowchart of the self-consistent cycle is shown in Figure 10[56]. Subsequent technical developments have further advanced FLEX. Esirgen and Bickers[57] extended it to multi-orbital lattice models, making it applicable to the study of multi-orbital superconductors. In summary, the two prominent features of FLEX are conservation and self-consistency. Compared to the straightforward calculations of the random phase approximation, the self-consistent solution process of FLEX requires greater computational power.

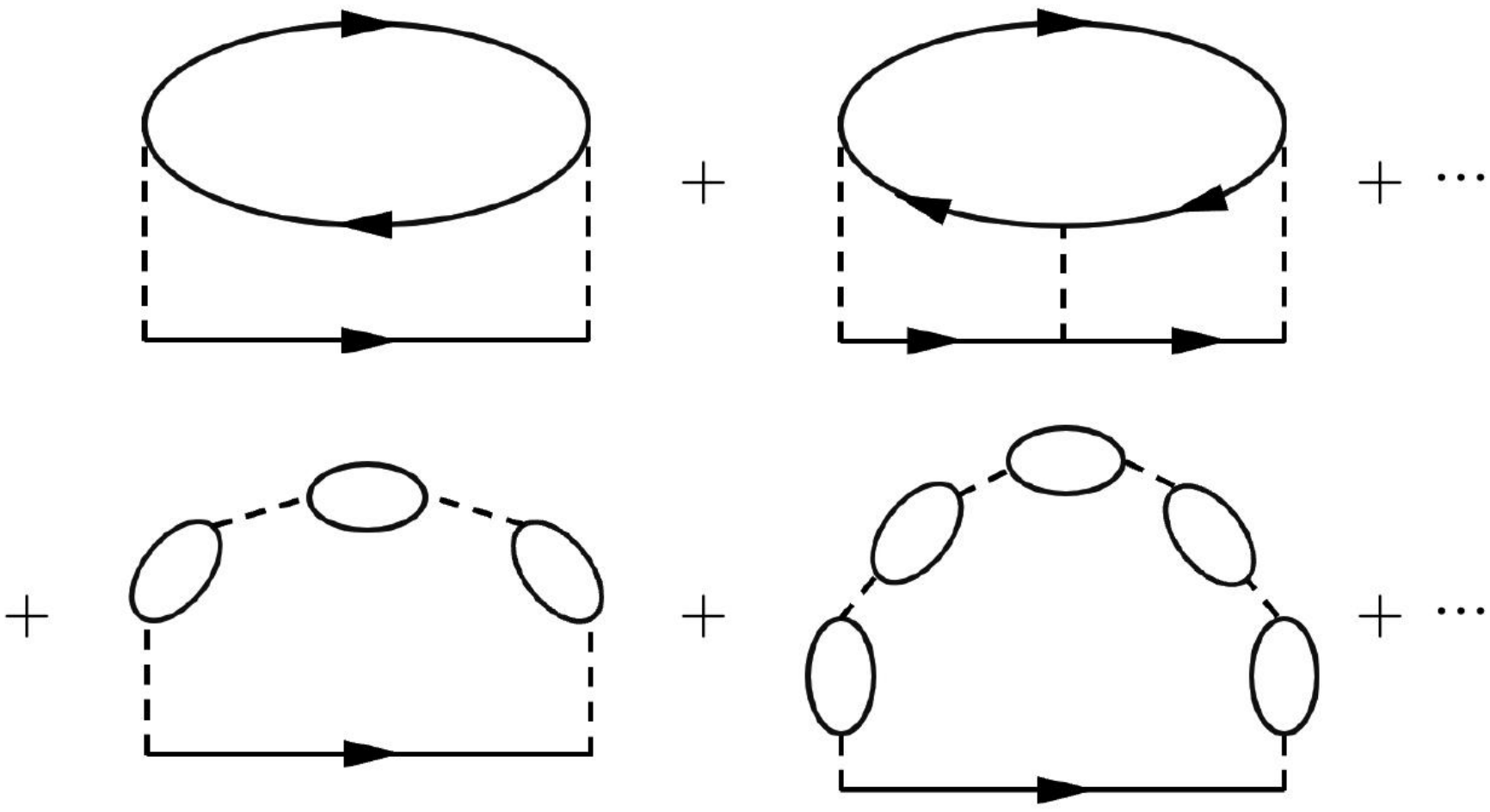

**Fig. 8 Feynman diagram of the self-energy at the FLEX level. The solid lines with arrows represent the "dressed" single-particle Green's function, and the dashed lines represent the "bare" interaction.**

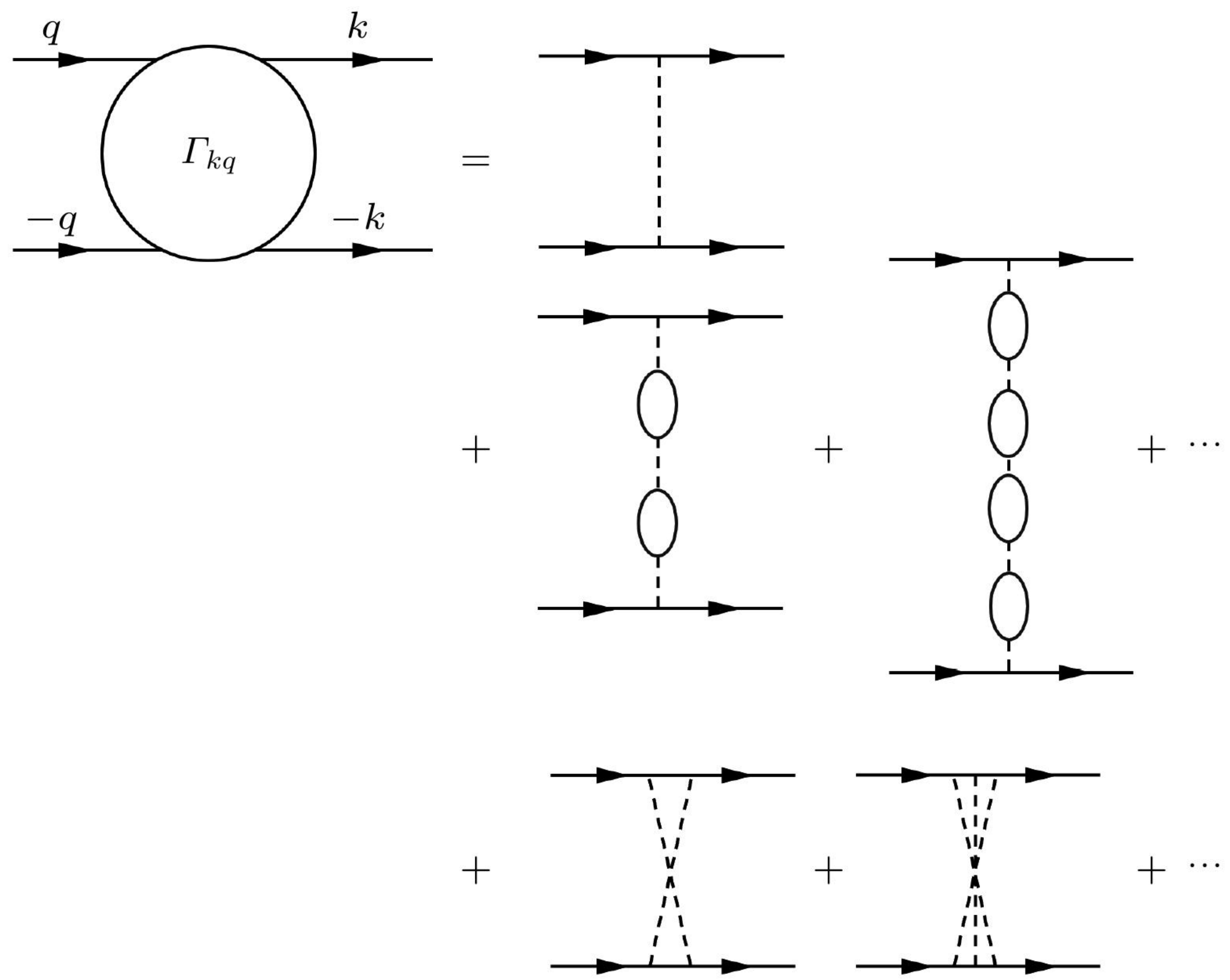

**Fig. 9 Feynman diagrams of the effective pairing interaction at the RPA level. The second line includes bubble contributions, and the third line includes contributions from (twist) ladders.**

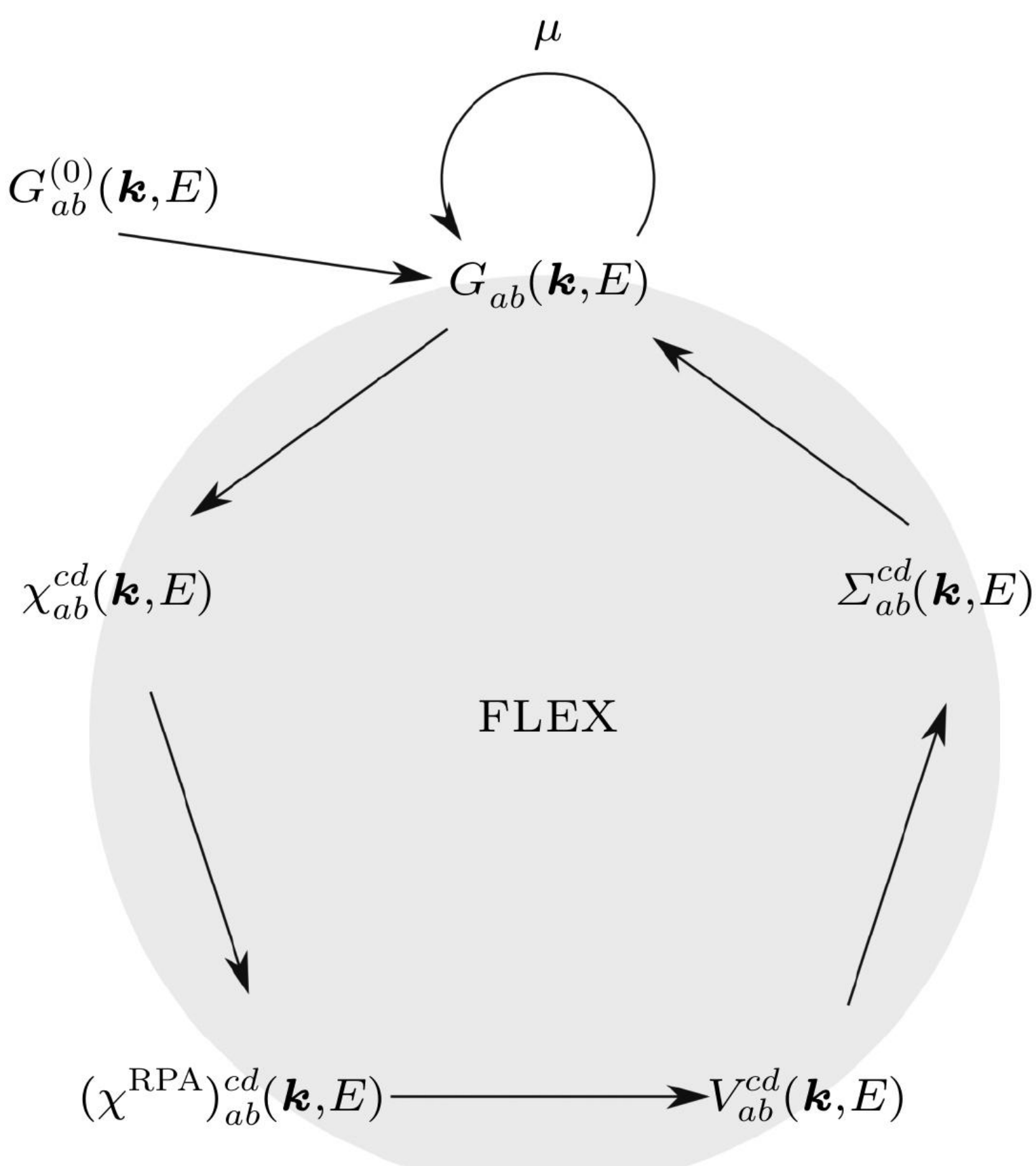

**Fig. 10 Flowchart for the self-consistent calculation of the Green's function and susceptibility in the FLEX loop[56].**

The fluctuation-exchange approximation was initially proposed to address unconventional superconductors in correlated electron systems. It has also exerted significant influence in the study of high-temperature superconductivity. Dahm and Tewordt[58] employed the fluctuation-exchange approximation to calculate the quasiparticle self-energy, energy gap, and spin susceptibility of the two-dimensional Hubbard model near half-filling. They focused on the momentum and frequency dependence of physical quantities. Their results revealed an instability toward spin density waves in the system. Furthermore, they observed an instability toward spin-singlet $d_{x^2-y^2}$-wave pairing in the vicinity of this instability. Kuroki et al.[59] used the fluctuation-exchange approximation to investigate the relationship between spin fluctuations and the Fermi surface in copper oxide high-temperature superconductors. By studying the single-band Hubbard model, they found that changes in the peak of spin fluctuations correlate with changes in the Fermi surface induced by hole doping. Ikeda et al.[60] examined how spin fluctuations and electron correlations in iron-based superconductors vary with doping. Using the fluctuation-exchange approximation to calculate the five-band Hubbard model, they discovered that appropriate hole doping positions low-energy spin excitations primarily near the $X$ point of the Brillouin zone. These excitations tend to become critical at low temperatures. Additionally, excessive hole doping results in weak momentum dependence for low-energy spin excitations. In contrast, electron doping introduces a gap in the spin excitation spectrum. Consequently, the corresponding nuclear magnetic resonance $1/T_1$ relaxation rate increases sharply after hole doping but is suppressed after electron doping. Wang Zidan et al.[61] utilized the fluctuation-exchange approximation to explore the link between spin fluctuations and unconventional superconducting pairing in iron-based high-temperature superconductors. They found that when interband antiferromagnetic spin fluctuations are significantly stronger than intraband ones, the system tends to form a spin-singlet extended s-wave superconducting state rather than a nodal $d_{xy}$-wave state. The reverse is also true. Moreover, effective interband coupling plays a crucial role in intraband pairing. Yu Shunli and Li Jianxin[62] systematically investigated the symmetry of spin fluctuations with different characteristics and superconducting gaps in iron-based superconductors. They performed calculations using the fluctuation-exchange approximation on multi-orbital tight-binding models. These models included Fermi surfaces with both electron and hole types, as well as those with only electron types. Their findings indicate that spin fluctuations are the common origin of superconductivity in iron-based high-temperature superconductors.

In theoretical studies of nickel-based 327 superconductors, the fluctuation-exchange approximation represents one of the prominent weak-coupling theories. Sakakibara et al.[63] applied the fluctuation-exchange approximation to study the four-orbital model of nickel-based 327 superconductors under high pressure. They found that the system favors $s_{\pm}$-wave pairing. This behavior resembles that of the bilayer Hubbard model[64]. High-temperature superconductivity originates from the nickel $d_{3z^2-r^2}$ orbital. Conversely, coupling with the $d_{x^2-y^2}$ orbital weakens superconductivity. Considering the similarity between the fluctuation exchange approximation and the random phase approximation, Heier et al.[65] employed a random phase approximation of the fluctuation exchange type. They calculated self-energies with momentum and frequency resolution. They combined this approach with density functional theory to investigate spin-fluctuation-induced pairing in nickel-based 327 superconductors under high pressure. A key feature of this method is that it does not require fitting band structures using tight-binding models. Instead, it directly uses electronic structures derived from first-principles calculations as the starting point. By numerically solving the gap function, they observed competition between $s_{\pm}$-wave and $d_{xy}$-wave pairing near the spin density wave instability. Li Jianxin et al.[66] extended the conventional bilayer two-orbital model for nickel-based 327 superconductors to further examine the impact of interlayer Coulomb interactions. Calculations using the fluctuation-exchange approximation revealed that when interlayer interactions are small, the system primarily exhibits $s_{\pm}$-wave pairing within the $d_{3z^2-r^2}$ orbital. As interlayer interactions increase, $s_{\pm}$-wave pairing is suppressed, while $d_{x^2-y^2}$-wave pairing is enhanced. In this regime, the resulting $d_{x^2-y^2}$-wave pairing is not intra-orbital. Instead, it involves pairing between interlayer $d_{x^2-y^2}$ and $d_{3z^2-r^2}$ orbitals. Further research indicates that the phase transition from $s_{\pm}$-wave to $d_{x^2-y^2}$-wave pairing is mainly driven by charge density fluctuations. Regarding the recently discovered nickel-based 327 superconductor films at ambient pressure, Ushio et al.[25] considered or ignored the $+U$ correction in band calculations. This approach yielded electronic structures with or without $\gamma$-pockets, respectively. However, calculations using the fluctuation-exchange approximation showed that the system exhibits $s_{\pm}$-wave pairing in both cases.

# 5 Functional Renormalization Group

The renormalization group concept related to condensed matter physics was first introduced by Kadanoff[67] as the "block spin" renormalization group while addressing the Ising model. The true power of the renormalization group method was demonstrated through Wilson's[68,69] landmark work on critical phenomena. This work yielded critical indices that surpass Landau's second-order phase transition theory and align more closely with experimental data[70]. Subsequently, numerical renormalization group methods were developed. Combined with computer simulations, these methods solved the Kondo problem in condensed matter physics[71]. They became the conceptual foundation for various subsequent numerical renormalization group techniques. These include the density matrix renormalization group[72,73], variational renormalization group[74], entanglement renormalization group[75,76], and tensor renormalization group[77-81]. As one of the numerical renormalization group methods, the functional renormalization group (FRG) derives its key equations from the study of effective actions in quantum field theory. Building on Polchinski's[82] proof of the renormalizability of four-dimensional $\phi^4$ theory using rigorous flow equations,

Wetterich[83] first obtained the rigorous functional flow equation for effective interactions. A key feature of this equation is its convenience for truncation to handle infrared divergence problems in quantum field theory. Morris[84] proposed reliable non-perturbative approximations based on the rigorous Polchinski flow equation and the Wetterich functional flow equation. Since then, the functional renormalization group method has been widely applied and developed across various fields of physics[85].

In the study of strongly correlated electron systems[86], the functional renormalization group was first applied to the Hubbard model on a two-dimensional square lattice related to copper-based high-temperature superconductivity. This application revealed phase transitions between metallic states and antiferromagnetic or BCS superconducting states[87-91]. The dominant pairing symmetry identified was d-wave pairing instability. Researchers also discovered novel phenomena such as Pomeranchuk instability[90], insulating spin liquid phases[91], and the coexistence of incommensurate magnetic order and superconductivity[92]. Wang et al.[93] used the functional renormalization group to investigate the pairing symmetry and mechanism in iron arsenide superconductors. By doping the five-band model and varying interaction strengths, they found that the dominant pairing symmetry is an extended s-wave pairing. In this state, the superconducting order parameters on electron and hole pockets have opposite signs. The superconducting instability is mainly driven by Josephson scattering between Fermi surfaces caused by antiferromagnetic correlations. By comparing one-loop functional renormalization group results for copper-based and iron-based superconductors, Wang et al.[94] found that superconducting pairing in both materials is driven by antiferromagnetic correlations. Additionally, Fermi surface distortion and orbital current order compete with this pairing. These ordering tendencies and antiferromagnetic correlations are characteristic of the same class of strongly correlated materials[95]. Platt et al.[96] used the functional renormalization group to study the phase diagram of microscopic models for iron arsenide superconductors. They established a connection between simplified two-band[97] and five-band[93] models. The primary pairing instability identified was extended s-wave pairing, driven by scattering between electron and hole pockets. Xiang et al.[98] also employed the functional renormalization group to study the pairing mechanism of monolayer FeSe film superconductors on $SrTiO_3$ substrates. They found that the screening effect of ferroelectric phonons in $SrTiO_3$ on Cooper pairs can significantly increase the energy scale of Cooper pairs. This effect can even alter the pairing symmetry. Since iron selenide superconductors differ from iron arsenide superconductors by having only electron pockets on the Fermi surface, Xiang et al.[99] used the functional renormalization group to find that s-wave superconducting pairing can still form on electron pockets. The pairing symmetry is $s^{++}$-wave. The superconducting pairing mechanism stems from the competition between two types of spin fluctuations. One is C-type spin fluctuations near collinear spin density waves, which induce attractive pairing scattering between electron pockets via Cooper pair excitations across virtual hole pockets. The other is G-type spin fluctuations near checkerboard spin density waves. These fluctuations cause repulsive pairing scattering, but hybridization splitting between electron pockets weakens their effect.

Compared to the random phase approximation and the fluctuation-exchange approximation, the functional renormalization group is the most powerful and persuasive weak-coupling theory in the theoretical study of nickel-based 327 superconductors. Yang et al.[17] used the functional renormalization group to investigate the pairing mechanism and symmetry in nickel-based 327 superconductors under high pressure. Through functional renormalization group calculations on the bilayer two-orbital Hubbard model, they found that in the weak-to-intermediate coupling regime, the primary superconducting pairing symmetry is $s_{\pm}$-wave pairing induced by spin fluctuations. In this state, the gap function has the same sign on the $\gamma$ and $\alpha$ pockets but the opposite sign on the $\beta$ pocket, as shown in Figure 11. In real space, the pairing primarily involves pairs with opposite phases on the same lattice site and orbital. Jiang et al.[100] studied the physical mechanism behind the monotonic decrease of the superconducting transition temperature with increasing pressure in nickel-based 327 superconductors. Density functional theory calculations revealed that as pressure increases from 14 GPa, the Fermi pockets remain almost unchanged, but the bandwidth increases. Additionally, interlayer hopping between $d_{3z^2-r^2}$ orbitals enhances. Further functional renormalization group calculations showed that as pressure increases, $s_{\pm}$-wave pairing induced by spin fluctuations weakens. This leads to a monotonic decrease in the superconducting transition temperature with increasing pressure. This finding contrasts sharply with the local magnetic moment picture of strong coupling. It provides strong evidence for the itinerant picture of d-electrons under weak coupling. Hu Jiangping and Wu Xianxin et al.[13] used the functional renormalization group to study nickel-based 327 superconductors under high pressure. They obtained the same pairing symmetry, namely $s_{\pm}$-wave pairing. They pointed out that interlayer and intralayer exchange couplings of the $d_{3z^2-r^2}$ orbital are crucial for high-temperature superconductivity. Hu Jiangping and Wu Xianxin et al.[101] further investigated the impact of electron-phonon coupling on superconductivity in nickel-based 327 superconductors under high pressure. Functional renormalization group calculations revealed that electron-phonon coupling alone is insufficient to generate superconductivity. They also found that different phonon modes couple to different nickel $d_{x^2-y^2}$ and $d_{3z^2-r^2}$ orbitals. This phenomenon is known as orbital-selective electron-phonon coupling. Specifically, interlayer electron-phonon coupling synergizes with electron correlations to promote interlayer pairing of the $d_{3z^2-r^2}$ orbital. In contrast, intralayer electron-phonon coupling only slightly affects superconductivity. This theory also predicts the oxygen isotope effect in nickel-based 327 superconductors under high pressure. Cao et al.[27] systematically studied the influence of substrate strain on nickel-based 327 superconductor films at ambient pressure. They simulated substrate effects using different in-plane compressions. Using DFT, they constructed a two-orbital tight-binding model. Functional renormalization group calculations revealed that $s_{\pm}$-wave pairing persists in the films.

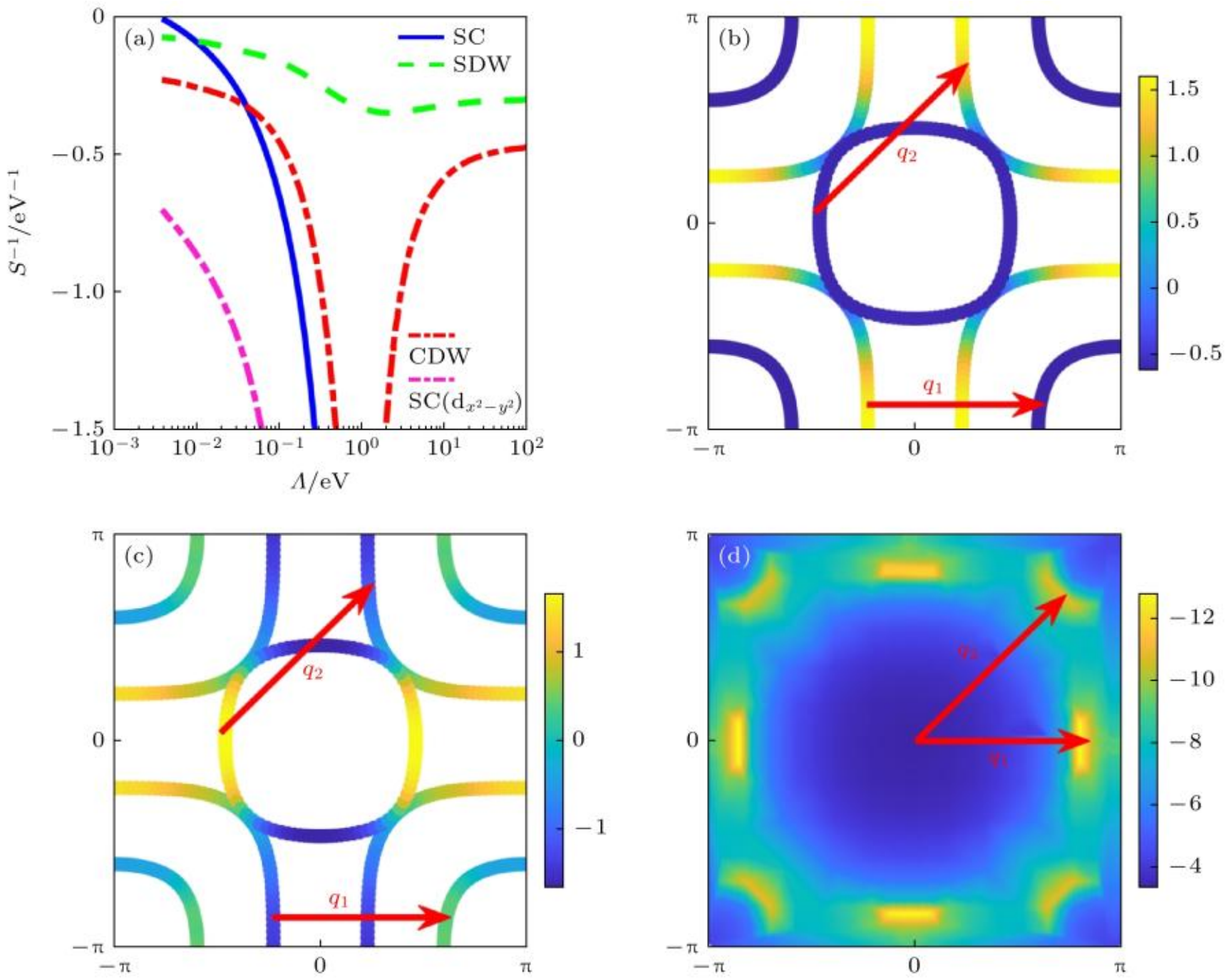

**Fig. 11 (a) FRG flow of the leading eigenvalues *S* versus the running energy scale *Λ*. Different colors represent the superconducting (SC), SDW, and CDW channels with parameters $U = 3$ eV and $J_{\mathrm{H}} = 0.3$ eV[17]. (b), (c) Gap functions for the leading s-wave and subleading $d_{x^2-y^2}$-wave on the Fermi surface, respectively[17]. (d) Renormalized interaction $V_{\mathrm{SDW}}(q)$ at the SC divergence, with two arrows indicating the dominant scattering momenta $q_{1,2}$[17].**

# 6 Conclusions and Discussion

The multi-orbital nature of the nickel-based 327 high-temperature superconducting bilayer system makes it suitable for description using a two-orbital Hubbard model.

This approach resembles that used for iron-based superconductors, although their detailed electronic structures differ. Such models have been extensively studied using various techniques, leading to several proposals for pairing mechanisms. These proposals can be broadly categorized into weak-coupling and strong-coupling approaches.

In weak-coupling theory, pairing arises from the exchange of spin and/or charge fluctuations enhanced by Fermi surface nesting. The topology of Fermi surfaces with multiple bands and favorable nesting conditions provides a reasonable basis for applying this theoretical framework. Techniques such as the RPA[18,23,44,45,102-105], fluctuation-exchange approximation (FLEX)[25,63,65], and FRG[13,17,27,106] mainly agree on $\mathrm{s}^{\pm}$-wave symmetry. In this state, interlayer coupling between $3\mathrm{d}_{z^2}$ orbitals plays a dominant role, highlighting the importance of the $\gamma$ pocket. The sensitivity of this pairing state to crystal field splitting[46] and interlayer interactions[66] has also been investigated. Furthermore, FRG studies on the pressure dependence of $T_\mathrm{c}$ show consistency with experimental observations[100]. Weak-coupling methods have also been applied to study superconductivity in RP bilayer thin films at ambient pressure[23]. Although most weak-coupling studies focus on an itinerant picture centered on $3\mathrm{d}_{z^2}$ orbitals and $\mathrm{s}^{\pm}$-wave states, some works have proposed alternative pairing symmetries or mechanisms[10,107-112].

In contrast to the itinerant picture, strong-coupling approaches assume that the bonding $3\mathrm{d}_{z^2}$ orbitals are nearly localized. This picture is motivated by the experimental observation of spin density waves in the normal state, which is a hallmark of strong electron correlations. Based on this premise, various models have been proposed, including several versions of the *t*-*J* model. Examples include the bilayer *t*-*J* model based on $3\mathrm{d}_{x^2-y^2}$ orbitals[113], the $t$-$t_\perp$-$J$ model for $3\mathrm{d}_{x^2-y^2}$ orbitals[11], the $t$-$t'$-$t''$-$J$ model for antisymmetric bands[114], the empty single-particle-double-particle *t*-*J* model[115], and the mixed-dimensional *t*-*J* model[116,117]. Other proposals involve multi-orbital models composed of interlayer antiferromagnetic coupling between $3\mathrm{d}_{z^2}$ orbitals, Hund's coupling between $3\mathrm{d}_{x^2-y^2}$ and $3\mathrm{d}_{z^2}$ orbitals, and *t*-*J*-like or Hubbard-like terms for $3\mathrm{d}_{x^2-y^2}$ orbitals or self-doped molecular Mott insulators[19,118-120]. These models have been studied using various techniques, such as slave-boson mean-field theory[114,113,121-124], renormalized mean-field theory (RMFT)[11,17,19,125-127], the slave-spin method[128,129], large-*N* expansion[130], density matrix renormalization group (DMRG)[118,124,131-137], tensor networks (TN)[136,138], static auxiliary field Monte Carlo[139], determinant quantum Monte Carlo (DQMC)[140], and variational quantum Monte Carlo (VQMC)[141]. Depending on the model and parameters, the predicted pairing symmetry ranges from s-wave to d-wave, and even $d + is$-wave.

A key issue in studying the pairing mechanism of nickel-based high-temperature superconductors is the precise role of the $3d_{z^2}$ orbital. Many theories, particularly those from a weak-coupling perspective, predict that the presence of the $\gamma$ pocket dominated by $3d_{z^2}$ enhances superconductivity, with pairing occurring primarily in the $3d_{z^2}$ orbitals. Conversely, other theories, especially those proposed within the strong-coupling framework, suggest that the $\gamma$ pocket is less relevant. Given that the $3d_{z^2}$ states are nearly localized, these theories argue that the superconducting order parameter should be dominated by the $3d_{x^2-y^2}$ orbitals. Therefore, experimentally determining the itinerant versus localized characteristics of the $3d_{z^2}$ orbital is crucial for resolving the superconducting mechanism. Theoretical research on nickel-based high-temperature superconductivity is still in its early stages and requires more precise experimental guidance. Further development of both weak-coupling and strong-coupling theories is necessary to advance the understanding of this phenomenon.

## Acknowledgement

This work was supported by the National Key Basic Research and Development Program of China (Grant No. 2024YFA1408101), the National Natural Science Foundation of China (Grant Nos. 12141402, 12334002, 12404171), and the Guangdong Project (Grant No. SZZX2401001).

[#] E-mail: miaojianjian@quantumsc.cn

[†] E-mail: chenwq@sustech.edu.cn